\documentclass[twocolumn,prd]{revtex4-1}

\usepackage{graphicx}
\usepackage{amssymb}
\usepackage{amsmath}

\newcommand{\abs}[1]{|#1|}

\newcommand{\diag}{\operatorname{diag}}

\begin{document}

\author{Kristjan Kannike}
 \email{kristjan.kannike@cern.ch}
\affiliation{National Institute of Chemical Physics and Biophysics, \\
R\"{a}vala 10, Tallinn 10143, Estonia}
\author{Kaius Loos}%
 \email{kaius.loos@gmail.com}
\affiliation{National Institute of Chemical Physics and Biophysics, \\
R\"{a}vala 10, Tallinn 10143, Estonia}
\author{Martti Raidal}%
 \email{martti.raidal@cern.ch}
\affiliation{National Institute of Chemical Physics and Biophysics, \\
R\"{a}vala 10, Tallinn 10143, Estonia}

\title{Gravitational Wave Signals of Pseudo-Goldstone Dark Matter in the $\mathbb{Z}_{3}$ Complex Singlet Model}

\begin{abstract}
We study pseudo-Goldstone dark matter in the $\mathbb{Z}_{3}$ complex scalar singlet model. Because the direct detection spin-independent cross section is suppressed, such dark matter is allowed in a large mass range. Unlike in the original model stabilized by a parity, due to the cubic coupling of the singlet the $\mathbb{Z}_{3}$ model can accommodate first-order phase transitions that give rise to a stochastic gravitational wave signal potentially observable in future space-based detectors. 
\end{abstract}

\maketitle

\section{Introduction}

One of the best candidates for the dark matter (DM) is a scalar singlet \cite{Silveira:1985rk,McDonald:1993ex}. The properties of singlet DM have been studied in detail \cite{Barger:2008jx,Burgess:2000yq,Cline:2013gha,Djouadi:2011aa} (see \cite{Athron:2017kgt,Arcadi:2019lka} for recent reviews; see also Refs. therein). The non-observation of dark matter by direct detection experiments \cite{Akerib:2016vxi,Aprile:2018dbl,Cui:2017nnn}, however, puts severe bounds on models of DM comprised of weakly interacting massive particles (WIMP), pushing the mass of singlet scalar DM -- except around the Higgs resonance -- over $1$~TeV. 

A way to suppress the direct detection cross section is to consider as the DM candidate a pseudo-Goldstone with derivative couplings to CP-even states \cite{Gross:2017dan} (the DM phenomenology of the imaginary part of the complex scalar singlet was first considered in \cite{Barger:2008jx,Chiang:2017nmu}, but only in the Higgs resonance region). Because the velocity of DM particles in the Galaxy is small, the prospective signal is suppressed by vanishing momentum transfer. This result remains practically unaffected when loop corrections to the direct detection cross section are taken into account \cite{Azevedo:2018exj,Ishiwata:2018sdi}. Despite that, it is possible for pseudo-Goldstone DM to show up at the LHC~\cite{Huitu:2018gbc,Alanne:2018zjm,Azevedo:2018oxv,Cline:2019okt}.

In this class of models, the global $U(1)$ symmetry is explicitly, but only softly broken into a discrete subgroup which is then broken spontaneously. In Ref.~\cite{Gross:2017dan}, the $U(1)$ group was explicitly broken into $\mathbb{Z}_{2}$ symmetry. We study the consequences of breaking $U(1)$ into $\mathbb{Z}_{3}$. The model admits two phases that produce a dark matter candidate. Firstly, the unbroken $\mathbb{Z}_{3}$ symmetry stabilizes $S$ as a dark matter candidate. Secondly, with the broken $\mathbb{Z}_{3}$ symmetry, the imaginary part  of $S$, denoted by $\chi$, is still stable due to the $S \to S^{\dagger}$ symmetry of the Lagrangian. The main difference between the $\mathbb{Z}_{2}$ and $\mathbb{Z}_{3}$ pseudo-Goldstone DM models is that the potential of the latter contains a cubic $S^{3}$ term. There are other possible cubic terms \cite{Jiang:2015cwa,Alves:2018oct,Alves:2018jsw}, but they do not respect the $\mathbb{Z}_{3}$ symmetry.

To this date, the $\mathbb{Z}_{3}$-symmetric complex singlet model has been studied in the unbroken phase. Originally, the model was proposed in the context of neutrino physics \cite{Ma:2007gq}. Detailed analysis of DM phenomenology was carried out in Ref.~\cite{Belanger:2012zr}.  Indirect detection of $\mathbb{Z}_{3}$ DM was considered in \cite{Arcadi:2017vis,Cai:2018imb}. The cubic coupling can contribute to $3 \to 2$ scattering for $\mathbb{Z}_{3}$ strongly interacting (SIMP) DM \cite{Hochberg:2014dra,Choi:2015bya,Daci:2015hca,Choi:2016tkj}. The effects of early kinetic decoupling were studied in \cite{Hektor:2019ote}. The $\mathbb{Z}_{3}$ symmetry has been considered as the remnant of a dark $U(1)$ local \cite{Ko:2014loa,Ko:2014nha,Guo:2015lxa} or global \cite{Bernal:2015bla} symmetry. In the unbroken phase, the direct detection cross section can also be suppressed if the DM relic density is determined by semi-annihilation processes \cite{Hambye:2008bq,Hambye:2009fg,Arina:2009uq,DEramo:2010ep,Belanger:2012vp,Belanger:2014bga}. However, the suppression is not as large as for pseudo-Goldstone DM.

The discovery of gravitational waves (GWs) by the LIGO experiment~\cite{Abbott:2016blz,Abbott:2016nmj} opened a new avenue to probe new physics. First-order phase transitions generate a stochastic GW background~\cite{Hogan:1984hx,Steinhardt:1981ct,Witten:1984rs} which may be discoverable in future space-based GW interferometers~\cite{Corbin:2005ny,Seoane:2013qna}.
While the SM Higgs phase transition is of second order~\cite{Kajantie:1996mn,Aoki:1999fi} and does not generate a GW signal, in models with extended scalar sector the first-order phase transition in the early Universe can become testable by observations. For a recent review on phase transitions and GWs, see Ref.~\cite{Mazumdar:2018dfl}.

GWs from beyond-the-SM physics with a scalar singlet have been studied in detail. These models admit a two-step phase transition that can be of the first order \cite{Espinosa:2011eu,Cline:2012hg,Espinosa:2011ax,Alanne:2014bra,Alanne:2016wtx,Tenkanen:2016idg,Zhou:2018zli,Cheng:2018ajh} and can potentially produce a measurable GW signal \cite{Kakizaki:2015wua,Hashino:2016xoj,Vaskonen:2016yiu,Huang:2016cjm,Artymowski:2016tme,Beniwal:2017eik,Chao:2017vrq,Bian:2017wfv,Huang:2018aja,Beniwal:2018hyi,Croon:2018erz,Dev:2019njv}. However, in the $\mathbb{Z}_{2}$ pseudo-Goldstone model, all phase transitions leading to the correct vacuum are of the second order~\cite{Kannike:2019wsn}, yielding no stochastic GW signal and excluding the additional potentially powerful experimental test of this class of the SM models. The phenomenology of phase transitions and GWs of the $\mathbb{Z}_{3}$ complex singlet model was studied in detail in Ref. \cite{Kang:2017mkl}. This work, however, disregarded the singlet as a DM candidate, because in the unbroken phase a sizeable GW signal is incompatible with correct relic density.

The goal of this paper is to study the nature of phase transitions and GW signals in the complex scalar singlet model in which a $U(1)$ symmetry softly broken into its $\mathbb{Z}_{3}$ subgroup. Just like in the $\mathbb{Z}_{2}$ case, the elastic scattering cross section pseudo-Goldstone DM with matter can be small enough for the DM to be well-hidden from direct detection. We find, however, that unlike in the $\mathbb{Z}_{2}$ case, strong first-order phase transitions can take place because the potential contains a cubic term. The resulting stochastic GW background is potentially discoverable in future space-based detectors such as LISA and BBO.

The paper is organized as follows. We introduce the model in Sec.~\ref{sec:model}. Various theoretical and experimental constraints are discussed in Sec.~\ref{sec:constraints}. DM relic density, phase transitions and the predictions for the direct detection and GW signals are treated in Sec.~\ref{sec:signals}. We conclude in Sec.~\ref{sec:conclusions}. The details on the effective potential and thermal corrections are relegated to Appendix~\ref{sec:counterterms}.

\section{$\mathbb{Z}_{3}$ Complex Singlet Model}
\label{sec:model}

The most general renormalizable scalar potential of the Higgs doublet $H$ and the complex singlet $S$, invariant under the $\mathbb{Z}_3$ transformation $H \to H$, $S \to e^{i 2 \pi/3} S$, is given by
\begin{equation}
\begin{split}
  V &= \mu_{H}^{2} \abs{H}^{2} + \lambda_{H} \abs{H}^{4} 
  + \mu_{S}^{2} \abs{S}^{2} + \lambda_{S} \abs{S}^{4}
   \\
  & + \lambda_{SH} \abs{S}^{2} \abs{H}^{2} + \frac{\mu_3}{2} (S^{3} + S^{\dagger 3}),
\end{split}
\label{eq:V:Z:3:singlet}
\end{equation}
where only the cubic $\mu_{3}$ term softly breaks a global $U(1)$ symmetry. Notice that the $\mathbb{Z}_{3}$ symmetry precludes any quartic couplings that would result in a \emph{hard} breaking of the $U(1)$. The potential Eq.~\eqref{eq:V:Z:3:singlet} -- just like in the original $\mathbb{Z}_{2}$ pseudo-Goldstone DM model -- has an additional discrete $\mathbb{Z}_{2}$ symmetry, $S \to S^{\dagger}$.

In the unitary gauge, we parameterize the fields as
\begin{equation}
  H = 
  \begin{pmatrix}
    0
    \\
    \frac{v + h}{\sqrt{2}}
  \end{pmatrix},
  \qquad
  S = \frac{v_{s} + s + i \chi}{2}.
\end{equation}
The $S \to S^{\dagger}$ is then equivalent to $\chi \to -\chi$, which makes $\chi$ stable even as the $\mathbb{Z}_{3}$ symmetry is broken. The model thus admits two different DM candidates: in the unbroken $\mathbb{Z}_{3}$ phase the complex singlet $S$ is a DM candidate, while in the broken $\mathbb{Z}_{3}$ phase it is its imaginary part, the pseudo-Goldstone $\chi$, which can be the DM. We concentrate on the latter case.

Without loss of generality, the vacuum expectation value (VEV) of $S$ can be taken to be real and positive, because the degenerate vacua where $\chi$ has a non-zero VEV are related to the real vacuum by $\mathbb{Z}_{3}$ transitions. This choice corresponds to a negative value of the parameter $\mu_{3}$.

The stationary point conditions are
\begin{align}
  h (2 \lambda_{H} h^2  + 2 \mu_{H}^{2}  + \lambda_{SH} s^2) &= 0, 
  \\
s (4 \lambda_{S} s^2  + 3 \sqrt{2} \mu_{3}  s + 4 \mu_{S}^{2}  + 2 \lambda_{SH} h^2) &= 0.
\end{align}  
We see that the model admits four types of extrema: the origin as $\mathcal{O} \equiv (0, 0)$ is fully symmetric; the vacuum  $\mathcal{H} \equiv (v_{h}, 0)$, with Higgs VEV only, spontaneously breaks the electroweak symmetry; the vacuum  $\mathcal{S} \equiv (0, v_{s})$, with $S$ VEV only, spontaneously breaks $\mathbb{Z}_{3}$; our vacuum $\mathcal{HS} \equiv (v_{h}, v_{s})$ breaks both symmetries.

The mass matrix of CP-even scalars in our $\mathcal{HS}$ vacuum is given by
\begin{equation}
  M^{2} = 
  \begin{pmatrix}
    2 \lambda_{H} v^{2} & \lambda_{SH} v v_{s}
    \\
    \lambda_{SH} v v_{s} & 2 \lambda_{S} v_{s}^{2} + \frac{3}{2 \sqrt{2}} \mu_{3} v_{s}
  \end{pmatrix}.
\label{eq:mass:matrix}
\end{equation}

The mass matrix \eqref{eq:mass:matrix} is diagonalised by an orthogonal matrix
\begin{equation}
  O = 
  \begin{pmatrix}
    \cos \theta & -\sin \theta 
    \\
    \sin \theta & \cos \theta
  \end{pmatrix}
\end{equation}
via $\diag (m_{1}^{2}, m_{2}^{2}) = O^{T} M^{2} O$. The mixing angle $\theta$ is given by 
\begin{equation}
   \tan 2 \theta = \frac{\lambda_{SH}  v v_{s}}{\lambda_{H} v^{2} + \lambda_{S} v_{s}^{2} - \frac{3}{4 \sqrt{2}} \mu_{3} v_{s}}.
\end{equation}
The mixing of the CP-even states $h$ and $s$ will yield two CP-even mass eigenstates $h_{1}$ and $h_{2}$
The mass of the pseudoscalar $\chi$ is, taking into account the extremum conditions,
\begin{equation}
  m_{\chi}^{2} = - \frac{9}{2 \sqrt{2}} \mu_{3} v_{s},
\end{equation}
which is proportional to $\mu_{3}$ as it explicitly breaks the $U(1)$ symmetry.

We express the potential parameters in terms of physical quantities in the zero-temperature vacuum, that is the masses $m_{1}^{2}$ and $m_{1}^{2}$ of real scalars, their mixing angle $\theta$, pseudoscalar mass $m_{\chi}^{2}$, and the VEVs $v_{h}$ and $v_{s}$:
\begin{align}
  \lambda_{H} &= \frac{ m_{1}^{2} + m_{2}^{2} + (m_{1}^{2} - m_{2}^{2}) \cos 2 \theta }{4 v_{h}^{2}},
  \\
  \lambda_{S} &= \frac{3(m_{1}^{2} + m_{2}^{2}) + 2 m_{\chi}^{2} + 3 (m_{2}^{2} - m_{1}^{2}) \cos 2 \theta}{12 v_{s}^{2}},
  \\
  \lambda_{SH} &= \frac{(m_{1}^{2} - m_{2}^{2}) \sin 2 \theta}{2 v_{s} v_{h}},
  \\
  \mu_{H}^{2} &= -\frac{1}{4} (m_{1}^{2} + m_{2}^{2}) + \frac{1}{4 v_{h}} (m_{2}^{2} - m_{1}^{2})
  \notag
  \\
  & \times (v_{h} \cos 2 \theta + v_{s} \sin 2 \theta),
  \\
  \mu_{S}^{2} &= -\frac{1}{4} (m_{1}^{2} + m_{2}^{2}) + \frac{1}{6} m_{\chi}^{2} + \frac{1}{4 v_{s}} 
  (m_{1}^{2} - m_{2}^{2}) 
  \notag
  \\
  & \times (v_{s} \cos 2 \theta - v_{h} \sin 2 \theta),
  \\
  \mu_{3} &= -\frac{2 \sqrt{2}}{9} \frac{m_{\chi}^{2}}{v_{s}}.
\end{align}
Both the Higgs doublet and the singlet will get a VEV, with the Higgs VEV given by $v_{h} = v = 246.22~\mathrm{GeV}$. We identify $h_{1}$ with the SM Higgs boson with mass 
$m_{1} = 125.09~\mathrm{GeV}$ \cite{Aad:2015zhl}.

\section{Theoretical and experimental constraints}
\label{sec:constraints}

We impose various theoretical and experimental constraints on the parameter space of the model.

First of all, the potential \eqref{eq:V:Z:3:singlet} is bounded from below if
\begin{equation}
  \lambda_{H} > 0, \quad \lambda_{S} > 0, \quad \lambda_{SH} + 2 \sqrt{\lambda_{H} \lambda_{S}} > 0.
\end{equation}

 Secondly, we require the couplings to be unitary and perturbative. 
The unitarity constraints in the $s \to \infty$ limit are given by
\begin{align}
  \abs{\lambda_{H}} \leqslant 4 \pi,  \quad\abs{\lambda_{S}} \leqslant 4 \pi, \quad  \abs{\lambda_{SH}} &\leqslant 8 \pi,
  \\
  \lvert 3 \lambda_{H} + 2 \lambda_{S}
  \notag
  \\
   \pm \sqrt{9 \lambda_{H}^2 - 12 \lambda_{H} \lambda_{S} + 4 \lambda_{S}^2 + 2 \lambda_{SH}^{2}} \rvert &\leqslant 8 \pi,
\end{align}
where the last condition, in the $\lambda_{SH} = 0$ limit, yields $\abs{\lambda_{H}} \leqslant \frac{4}{3} \pi$ and $\abs{\lambda_{S}} \leqslant 2 \pi$.
We also calculate unitarity constraints at finite energy with the help of the latest version \cite{Goodsell:2018tti} of the SARAH package \cite{Staub:2009bi,Staub:2010jh,Staub:2012pb,Staub:2013tta}. Scattering at finite energy allows to set a bound on the cubic coupling $\mu_{3}$.

To ensure validity of perturbation theory, loop corrections to couplings should be smaller than their tree-level values. The model is perturbative \cite{Lerner:2009xg} if $\abs{\lambda_{H}} \leqslant \frac{2}{3} \pi$, $\abs{\lambda_{S}} \leqslant \pi$ and $\abs{\lambda_{SH}} \leqslant 4 \pi$.

We require that our $\mathcal{HS}$ vacuum be the global one: this implies that
\begin{equation}
  m_{\chi}^{2} < \frac{9 m_{1}^{2} m_{2}^{2}}{m_{1}^{2} \cos^2 \theta + m_{2}^{2} \sin^{2} \theta},
\end{equation}
which for $\sin \theta \approx 0$ is approximated by $m_{\chi} \lesssim 3 m_{2}$. This bound appears to be stronger than the constraint on the cubic coupling from unitarity at finite-energy scattering.

If the mass of $x \equiv \chi \text{ or }h_2$ is less than $m_h/2$, then the Higgs invisible decay width into this particle is given by
\begin{equation}
  \Gamma_{h \to x x} = \frac{g_{hxx}}{8 \pi} \sqrt{1 - 4 \frac{m_x^2}{m_h^2}},
\end{equation}
with
\begin{align}
  g_{h1 \chi \chi} &= \frac{m_h^2 + m_\chi^2}{v_S},
  \\
  g_{h1 h_2 h_2} &= \frac{1}{v v_S} \left[ \left(\frac{1}{2} m_h^2 + m_2^2\right) (v \cos \theta + v_S \sin \theta) \right.
  \notag
  \\
  & \left. 
  + \frac{1}{6} v m_\chi^2 \cos \theta \right] \sin 2 \theta.
\end{align}
The invisible Higgs branching ratio is then given by
\begin{equation}
\text{BR}_\text{inv} = \frac{\Gamma_{h \to \chi\chi} + \Gamma_{h \to h_2 h_2}}{\Gamma_{h_1 \to \text{SM}} + \Gamma_{h \to \chi\chi} + \Gamma_{h \to h_2 h_2}},
\end{equation}
which is constrained to be below about $0.24$ at $95\%$ confidence level \cite{Khachatryan:2016whc,ATLAS-CONF-2018-031} by direct measurements and below about $0.17$ by statistical fits of all Higgs couplings \cite{Giardino:2013bma,Belanger:2013xza}. In extended Higgs models~\cite{Raidal:2011xk} the mixing phenomenology can be more complicated.

The mixing angle between $h$ and $s$ is constrained from the measurements of the Higgs couplings at the LHC to $\abs{\sin \theta} \leq 0.37$ for $m_{2} \gtrsim m_{h}$ and $\abs{\sin \theta} \leq 0.5$ for $m_{2} \lesssim m_{h}$~\cite{Ilnicka:2018def}.

Last, but not least, we require that the relic density of $\chi$ be equal to the value $\Omega h^{2} = 0.120 \pm 0.001$ from recent Planck data~\cite{Aghanim:2018eyx}.

\begin{figure*}[t]
\begin{center}
  \includegraphics{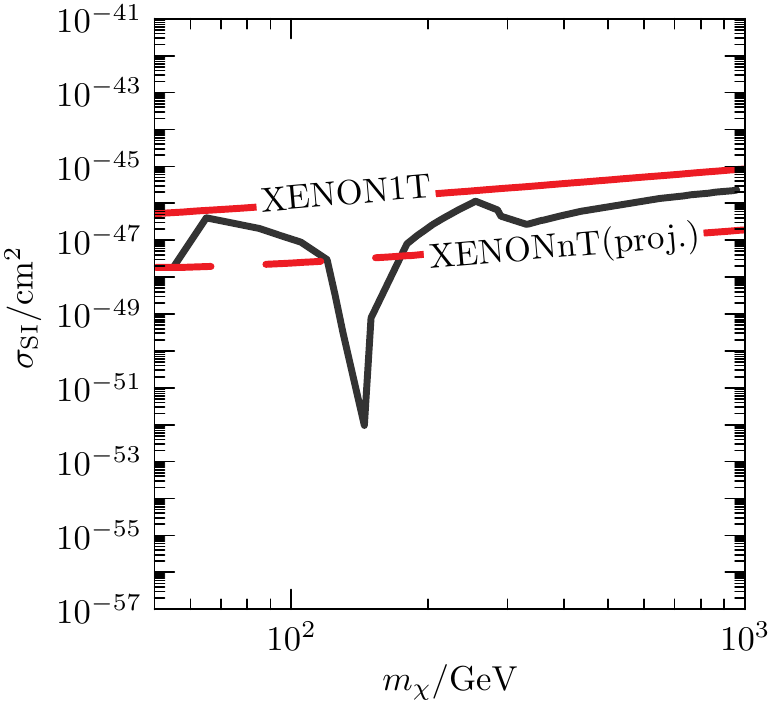}
  \qquad
  \includegraphics{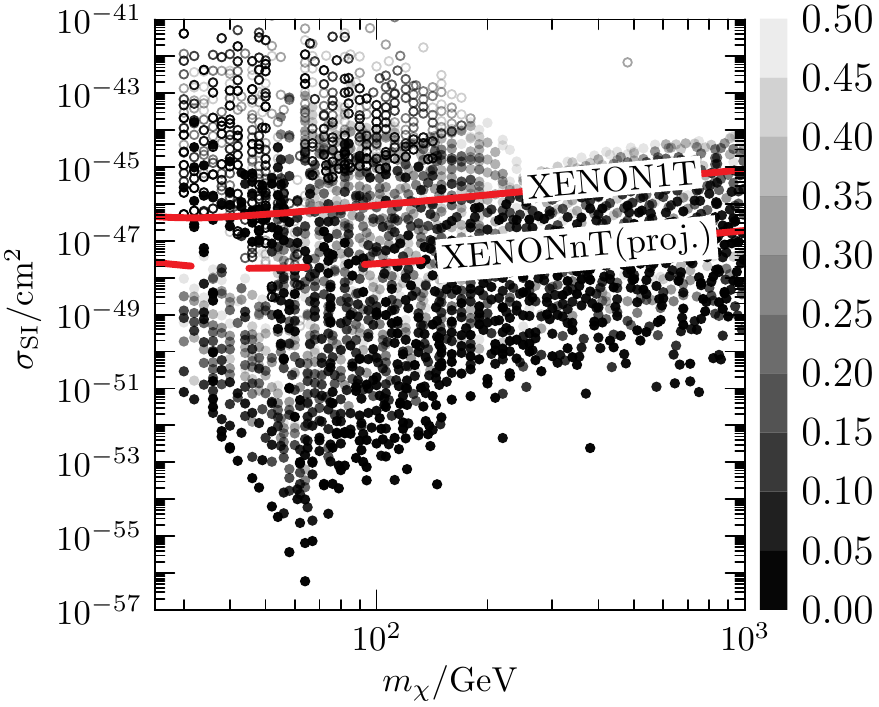}
\caption{Left panel: Spin-independent direct detection cross section $\sigma_{\text{SI}}$ as a function of the DM mass $m_{\chi}$ for fixed $m_{2} = 300$~GeV and $\sin \theta = 0.05$.  The bound from the XENON1T experiment is shown in red and the predicted sensitivity of the XENONnT experiment in dashed red. 
Right panel: Scatter plot of the results in $(m_{\chi}, \sigma_{\mathrm{SI}})$ plane in the grey-scale representing the dependence on $\abs{\sin \theta}$.
The upper bound in $\sigma_{\mathrm{SI}}$ is given by the requirement that the $(v_{h}, v_{s})$ minimum be the global one. In unfilled points, the Higgs invisible width surpasses the experimental limit. }
\label{fig:direct:detection}
\end{center}
\end{figure*}

\section{Direct detection and gravitational wave signals}
\label{sec:signals}

For the scan of the parameter space we choose $m_{\chi}$, $m_{2}$ and $\sin \theta$ as the free parameters, while $v_{s}$ is used to fit the DM relic density. We generate the free parameter values in the following ranges: $m_{\chi} \in [25, 1000]$~GeV, $m_{2} \in [25, 4000]$~GeV and $\sin \theta \in [-0.5, 0.5]$~GeV. We use the micrOMEGAs package \cite{Belanger:2018mqt} to fit the DM relic density and CosmoTransitions \cite{Wainwright:2011kj} for the calculation of phase transitions and the Euclidean action to determine tunnelling rates. The stochastic gravitational wave signal was calculated using the formulae of Ref. \cite{Caprini:2015zlo} for the non-runaway case.

The micrOMEGAs is also used to compute predictions for direct detection signals, to which we add the tiny loop correction. Unlike in the $\mathbb{Z}_{2}$ pseudo-Goldstone DM model, the tree-level direct detection DM amplitude contains a term that does not vanish at zero momentum transfer, so
\begin{equation}
  \mathcal{A}_{\text{dd}}(t \approx 0) \propto \lambda_{SH} \mu_{3}.
\label{eq:cancel}
\end{equation}
For a large part of the parameter space, however, this term is small enough, which allows one to explain the negative experimental results from DM direct detection experiments for a wide range of pseudo-Goldstone DM mass.

In the high temperature approximation \eqref{eq:high:T:approx}, the mass terms acquire thermal corrections:
\begin{equation}
    \mu_{H}^{2}(T) = \mu_{H}^{2} + c_{H} T^{2},
    \quad
    \mu_{S}^{2}(T) = \mu_{H}^{2} + c_{S} T^{2},
\end{equation}
with the coefficients
\begin{align}
   c_{H} &= \frac{1}{48} (9 g^{2} + 3 g^{\prime 2} + 12 y_{t}^{2} + 24 \lambda_{H} + 4 \lambda_{SH}),
   \\
   c_{S} &= \frac{1}{6} (2 \lambda_{S} + \lambda_{SH}).
\end{align}
For numerical calculations of gravitational wave signals, exact expressions are used.

We present in Fig.~\ref{fig:direct:detection} the results of our scans for the direct detection signal. In the left panel we present for illustration the behavior of spin-independent direct detection cross section $\sigma_{\text{SI}}$ as a function of DM mass $m_{\chi}$ for fixed values of $m_{2} = 300$~GeV and $\sin \theta = 0.05$. Red lines show the current limit from the XENON1T experiment \cite{Aprile:2018dbl} and the future bound from the projected XENONnT experiment \cite{Ni2017}. While the usual $\mathbb{Z}_{2}$ scalar singlet DM is excluded below $O(\text{TeV})$ masses, except for the Higgs resonance region, the suppression of the pseudo-Goldstone cross section in this model allows for lighter DM candidates. The two pronounced dips in the cross section are given by the Higgs resonance at $m_{h}/2$ and the $h_{2}$ resonance at $m_{2}/2$.

\begin{figure*}[t]
\begin{center}
  \includegraphics{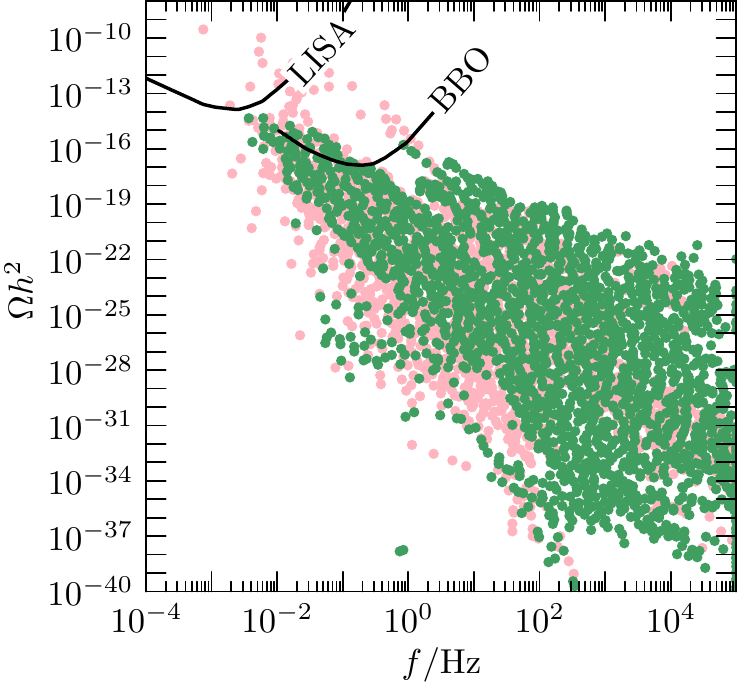}\qquad\qquad
  \includegraphics{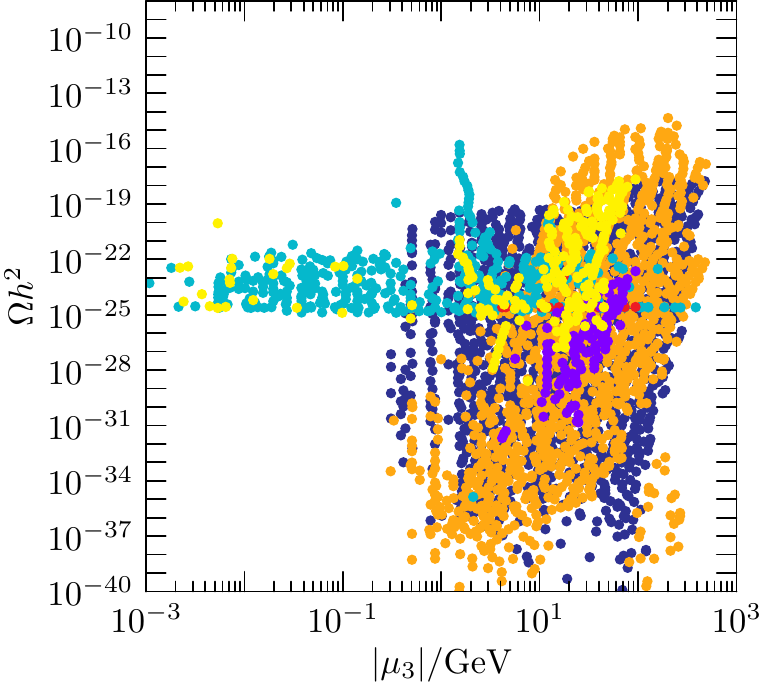}
\caption{Left panel: Scatter plot for the peak power of GWs vs. frequency together with the sensitivity curves of the LISA and BBO experiments. Light red points are forbidden and darker green points are allowed by the results of the XENON1T direct detection searches. Right panel: The dependence of the peak power of GWs on the value of cubic coupling. The color code shows the nature of the phase transition that gives the strongest signal: $\mathcal{O} \to \mathcal{S}$ in orange, $\mathcal{S} \to \mathcal{S}$ in dark blue, $\mathcal{S} \to \mathcal{HS}$ in cyan, $\mathcal{HS} \to \mathcal{HS}$ in purple, $\mathcal{H} \to \mathcal{HS}$ in yellow, $\mathcal{O} \to \mathcal{H}$ in red.}
\label{fig:gw}
\end{center}
\end{figure*}

The direct detection signal varies over a wide range and is roughly proportional to the cubic coupling $\mu_{3}$. Because the amplitude is proportional to $\sin \theta \cos \theta$, larger mixing angles also produce larger direct detection signal. The right panel in Fig.~\ref{fig:direct:detection}  presents the results of the whole scan, coded in grey-scale by the mixing angle $\abs{\sin \theta}$; empty points are excluded by the constraint on the Higgs invisible branching ratio. Note that the $\text{BR}_{\text{inv}}$ can exclude points with $m_{\chi} > m_{h}/2$, because the Higgs boson can decay invisibly also to two $h_{2}$ if it is light enough. Early kinetic decoupling \cite{Binder:2017rgn, Duch:2017nbe} may additionally enhance $\text{BR}_{\text{inv}}$ several times \cite{Hektor:2019ote}, but in practice this would not change the parameter space of GW signals. In particular, all our points with a potentially measurable GW signal have $m_{\chi}, m_{2} >m_{h}/2$. The upper bound on $\sigma_{\text{SI}}$ arises from the requirement the $\mathcal{HS}$ vacuum be global, which bound is stronger than that from unitarity that also constrains the points from above.

In Fig.~\ref{fig:gw} we present the predicted stochastic GW signal together with the predicted sensitivity curves of the future LISA and the BBO satellite experiments. To avoid cluttering the plot, we only show the peak power of each GW spectrum. In the left panel, light red points are excluded by the XENON1T direct detection constraints, while the darker green points are still allowed by direct detection; all points not satisfying the other constraints have been excluded. In the right panel, we depict the dependence of the GW signal on the cubic coupling $\mu_{3}$. The color code shows the nature of the phase transition that yields the strongest signal in a sequence of phase transitions. The largest signal is produced by the $\mathcal{O} \to \mathcal{S}$ transitions, enhanced by a sizable cubic coupling. Because $m_{\chi} \propto \mu_{3}$, there is a similar dependence on $m_{\chi}/m_{2}$. The strongest signals can be produced at roughly $m_{\chi} \approx 2.3 m_{2}$.

\section{Conclusions}
\label{sec:conclusions}

We have studied the $\mathbb{Z}_{3}$ complex scalar singlet DM model, where only the cubic coupling of the singlet explicitly breaks a global $U(1)$ symmetry. The model has two phases with stable DM. In the phase where the $\mathbb{Z}_{3}$ is spontaneously broken, the residual CP-like $\mathbb{Z}_{2}$ symmetry stabilizes the imaginary part of the complex singlet $S$ as a pseudo-Goldstone DM candidate. The DM direct detection cross section can be considerably suppressed by the small momentum transfer or the resonance at half the mass of the heavy singlet-like particle.

While we may be unable to discover the pseudo-Goldstone DM via direct detection, it may be testable by other means. In this model a stochastic gravitational wave background can arise from the first-order phase transitions due to the presence of a cubic coupling for the singlet. The strongest signals, which can potentially be observed by the future BBO experiment, are produced in the $\mathcal{O} \to \mathcal{S}$ phase transition (if present) at $m_{\chi} \approx 2.3 m_{2}$. The strength of the gravitational wave signal is anti-correlated with a small mixing angle and is, therefore, greater where the direct detection cross section is smaller (at fixed cubic coupling).

\section*{Acknowledgements}

We would like to thank Matti Heikinheimo and Christian Gross for useful comments. This work was supported by the Estonian Research Council grant PRG434, the grant IUT23-6 of the Estonian Ministry of Education and Research, and by the European Union through the ERDF Centre of Excellence program project TK133.

\appendix

\section{Thermal Effective Potential}
\label{sec:counterterms}

At one-loop level, the quantum corrections to the scalar potential in the $\overline{\text{MS}}$ renormalisation scheme are given by
\begin{equation}
  \Delta V = \sum_{i} \frac{1}{64 \pi^{2}} n_{i} m^{4}_{i} \left( \ln \frac{m_{i}^{2}}{\mu^{2}} - c_{i} \right),
\end{equation}
where $n_{i}$ are the degrees of freedom of the $i$-th field, $m_{i}$ are field-dependent masses and the constants $c_{i} = \frac{3}{2}$ for scalars and fermions and $c_{i} = \frac{5}{6}$ for vector bosons.
The masses and degrees of freedom $n_{i}$ of the fields are given in Table~\ref{tab:field:dependent:masses}. 
The field-dependent masses of $h$ and $s$ are given by the eigenvalues $m^{2}_{1,2}$ of the mass matrix of the CP-even eigenstates with elements given by
\begin{align}
  (m_{R}^{2})_{11} &= \mu_{H}^{2} + 3 h^{2} \lambda_{H}+ \frac{1}{2} \lambda_{SH} s^2,
  \\ 
  (m_{R}^{2})_{12} &= \lambda_{SH} h s,
    \\
  (m_{R}^{2})_{22} &= \mu_{S}^{2} + 3 \lambda_{S} s^2 + \frac{3}{2} \sqrt{2} \mu_{3} s 
     + \frac{1}{2} h^{2} \lambda_{SH}.
\end{align}
We neglect the contributions of the Goldstone bosons $G^{0}$ and $G^{\pm}$.
To calculate the effective potential in case of negative field-dependent masses, we substitute $\ln m_{i}^{2} \to \ln \abs{m_{i}^{2}}$, which is equivalent to analytical continuation \cite{Bobrowski:2014dla}. We set the renormalisation scale to $\mu = M_{t}$.

We add a counter-term potential,
\begin{equation}	
\begin{split}
  \delta V &= \delta\mu_{H}^{2} \abs{H}^{2} + \delta\lambda_{H} \abs{H}^{4} 
  + \delta\mu_{S}^{2} \abs{S}^{2} + \delta\lambda_{S} \abs{S}^{4}
  \\
  &+ \delta\lambda_{SH} \abs{S}^{2} \abs{H}^{2}  + \frac{\delta\mu_3}{2} (S^{3} + S^{\dagger 3}) + \delta V_{0},
\end{split}
\label{eq:delta:V:Z:3:singlet}
\end{equation}
in order to fix the VEVs and the mass matrix in our $\mathcal{HS}$ minimum to their tree-level values.

The thermal corrections to the potential are given by
\begin{equation}
  V_{T} = \frac{T^{4}}{2 \pi^{2}} \sum_{i} n_{i} J_{\mp} \left( \frac{m_{i}}{T} \right),
\end{equation}
where
\begin{equation}
  J_{\mp} = \pm \int_{0}^{\infty} dy y^{2} \ln \left[ 1\mp \exp \left( -\sqrt{y^{2} + x^{2}} \right) \right],
\end{equation}
with the $-$ sign applied to bosons and the $+$ sign to fermions. In the high-temperature limit $m/T \ll 1$, the thermal contributions are given by 
\begin{equation}
  V_{T}(T) = \frac{T^{2}}{24} \sum_{i} n_{i} m_{i}^{2}.
\label{eq:high:T:approx}
\end{equation}

The full thermally corrected effective potential is then
\begin{equation}
  V^{(1)} = V + \Delta V + \delta V + V_{T}.
\end{equation}

The counter-term potential $\delta V$ is chosen such as to keep quantum corrections to the masses, to mixing between $h$ and $s$ and to the VEVs zero. The counter-terms are given by
\begin{align}
  \delta \lambda_{H} &= \frac{1}{2 v^{3}} (\partial_{h} \Delta V - v \partial_{h}^{2} \Delta V),
  \\
  \delta \lambda_{S} &= \frac{1}{6 v_{S}^{3}} (4 \partial_{s} \Delta V 
  - v_{S} \partial_{\chi}^{2} \Delta V -3 v_{S} \partial_{s}^{2} \Delta V),
  \\
  \delta \lambda_{SH} &= -\frac{1}{v v_{S}} \partial_{h} \partial_{s} \Delta V,
  \\
  \delta \mu_{H}^{2} &= \frac{1}{2 v} (- 3 \partial_{h} \Delta V + v_{s} \partial_{h} \partial_{s} \Delta V
  + v \partial_{h}^{2} \Delta V),
  \\
  \delta \mu_{S}^{2} &= -\frac{1}{6 v_{S}} (v_{S} \partial_{\chi}^{2} \Delta V + 8 \partial_{s} \Delta V
  - 3 v_{S} \partial_{s}^{2} \Delta V 
  \notag
  \\
  &- 3v \partial_{h} \partial_{s} \Delta V),
  \\
  \delta \mu_{3} &= \frac{2 \sqrt{2}}{9 v_{S}^{2}} (v_{S} \partial_{\chi}^{2} \Delta V - \partial_{s} \Delta V),
\end{align}
where we take $h = v$, $s = v_{S}$, $\chi = 0$ after taking the derivatives.

\begin{table}[tb]
\caption{Field-dependent masses and the numbers of degrees of freedom.}
\begin{center}
\begin{tabular}{ccc}
  Field $i$ & $m_{i}^{2}$ & $n_{i}$ \\
  \hline
  $h$ & $m^{2}_{1}$ & $1$ \\
  $s$ & $m^{2}_{2}$ & $1$ \\
  $\chi$ & $\mu_{S}^{2} + \lambda_{S} s^2 - \frac{3}{\sqrt{2}} \mu_{3} s 
  + \frac{1}{2} \lambda_{SH} h^{2}  $ & $1$ \\
  $Z^{0}$ & $\frac{1}{4} (g^{2} + g^{\prime 2}) h^{2}$ & $3$ \\
  $W^{\pm}$ & $\frac{1}{4} g^{2} h^{2}$ & $6$ \\
  $t$ & $\frac{1}{2} y_{t} h^{2}$ & $-12$ \\
\end{tabular}
\end{center}
\label{tab:field:dependent:masses}
\end{table}

\bibliography{U1sbZ3cxSM}

\end{document}